\documentclass[12pt,aps,showpacs,showkeys]{revtex4}
\usepackage{hyperref}
\usepackage{amsmath}
\usepackage{bm}
\usepackage{graphicx}

\begin{document}

\title{Delay-induced synchronization phenomena in an array of globally
  coupled logistic maps}

\author{A. C.  Mart\'{\i} and C. Masoller} \affiliation{Instituto de
  F\'{\i}sica, Facultad de Ciencias, Universidad de la Repu\'blica,
  Igu\'a 4225, 11400 Montevideo, Uruguay}

\date{\today}

\begin{abstract}
  
  We study the synchronization of a linear array of globally coupled
  identical logistic maps. We consider a time-delayed coupling that
  takes into account the finite velocity of propagation of the
  interactions. We find globally synchronized states in which the
  elements of the array evolve along a periodic orbit of the uncoupled
  map, while the spatial correlation along the array is such that an
  individual map sees all other maps in his present, current, state.
  For values of the nonlinear parameter such that the uncoupled maps
  are chaotic, time-delayed mutual coupling suppress the chaotic
  behavior by stabilizing a periodic orbit which is unstable for the
  uncoupled maps. The stability analysis of the synchronized state
  allows us to calculate the range of the coupling strength in which
  global synchronization can be obtained.

\end{abstract}
\keywords{ Synchronization; coupled map arrays; time delays; logistic map}
\pacs{05.45.Xt, 05.65.+b, 05.45.Ra}
\maketitle 

\section {Introduction}

Coupled oscillator models are widely used to model complex dynamics in
non-equilibrium extended systems, and their synchronization has
attracted a lot of attention in recent years \cite{book}. In studies
of coupled ensembles of nonlinear oscillators, different situations
have been considered (identical or non-identical units, periodic or
chaotic single-unit behavior, local or global coupling), and a rich
variety of synchronization phenomena has been found (for a recent
review, see \cite{review}).

In the field of coupled map lattices, the paradigmatic model,
originally introduced by Kaneko \cite{Kaneko_1989,Kaneko_1990}, is the
ensemble of $N$ logistic maps with mean field global coupling:
\begin{equation} 
\label{mapa0}
x_i(t+1)= (1-\epsilon) f[x_i(t)] + \frac {\epsilon}N \sum_{j=1}^N f[x_j(t)], 
\end{equation}
$i\in [1,N]$, $f(x)=ax(1-x)$,
and $\epsilon$ is the coupling strength. For relatively large coupling
global (full) synchronization occurs: the array synchronizes on the
manifold $x_1=...=x_N$, where the dynamics of an element is generated
by the uncoupled map. For weaker coupling cluster (or partial)
synchronization occurs: the array splits into $K$ clusters of
$N_1,...,N_K$ elements mutually synchronized
\cite{Pikovsky_PRL,Pikovsky_PhysicaD}.

A characteristic of many biological and physical systems is
time-delayed coupling in the interaction among many units. In the case
of globally coupled units, the introduction of time delays makes the
spatial coordinates of an element relevant in spite of the infinite
range of the mean-field interaction. This situation was considered in
Ref.\cite{Zanette_2000} for one-dimensional arrays of coupled phase
oscillators. It was shown that in the limit of short delays the
ensemble approaches a state of frequency synchronization, and that
this state might develop a spatial nontrivial distribution of phases.
In two-dimensional arrays, distance-dependent time delays induce a
variety of patters including traveling rolls, steady patterns,
spirals, and targets \cite{PRL_2002}.

Here we study retardation effects in a linear array of logistic maps:
\begin{equation} 
\label{mapa}
x_i(t+1)= (1-\epsilon) f[x_i(t)] + 
\frac {\epsilon} N\sum_{j=1}^N f[x_j(t-\tau_{ij})],
\end{equation}
where $\tau_{ij} = k |i-j|$ is proportional to the distance between
the $i$th and $j$th maps and $k$ is the inverse of the velocity of the
signal that travels through the array. In a previous work
\cite{Physica_A} we considered the case in which the uncoupled maps
evolve in a periodic orbit of period 2 (when $3\leq a \leq 1 +
\sqrt{6}$).  We found that for weak coupling the array divides into
clusters, and the behavior of the individual elements within each
cluster depend on the delay times. For strong enough coupling global
synchronization occurs, where the dynamics of an element is periodic
of period 2, generated by the uncoupled logistic map. The spatial
correlation of the elements along the array is such that if $k$ is
even, at time $t$ all elements are in the same state, while if $k$ is
odd, at time $t$ neighboring elements are in different states.  In
both cases an individual map sees all other maps in his present,
current, state.

In this paper we extend the previous study and consider that the
uncoupled maps can be either periodic or chaotic (i.e., $3\leq a\leq
4$). We find that for adequate coupling strength and time delay,
global synchronization occurs. In the globally synchronized state all
elements evolve along a periodic orbit of the uncoupled logistic map.
Remarkably, this orbit might be unstable for the uncoupled maps. In
particular, when the uncoupled maps are chaotic, time-delayed coupling
might suppress chaos, stabilizing an unstable periodic orbit.  For
small arrays we study the stability of the globally synchronized
solution and calculate the minimum coupling strength above which the
unstable orbit of the uncoupled maps becomes stable for the
time-delayed coupled maps. The numerical simulations are in excellent
agreement with the stability analysis.

This paper is organized as follows. In Sec. II we analyze the
existence and the stability of the globally synchronized state. In
Sec. III we present results of the numerical simulations and the
stability analysis. Finally, in Sec. IV we present a summary and the
conclusions.

\section{Globally synchronized solutions}

A special class of solutions of Eq. (\ref{mapa}) is characterized by
the fact that, for all pairs $i$, $j$, the signal received by map $i$
at each time corresponds to a {\it delayed} state of map $j$ that
coincides with the {\it present} state of map $i$:
\begin{equation}
x_j(t-\tau_{ij})=x_i(t).
\end{equation}
Thus, each element ``perceives'' the array as being fully synchronized, in
spite of the fact that the simultaneous states of different elements might
not coincide. In these globally synchronized solutions each element
evolves along a limit cycle of period $P$ of the uncoupled logistic map with a
given phase, such that we can write
\begin{equation}
x_i(t)=x_0(t+\phi_i)
\end{equation}
with $x_0(t)$ a particular realization of the limit cycle, used as a
reference orbit. The condition for this solution to satisfy the
evolution equation is
\begin{equation} \label{fase}
\phi_i - \phi_j +m_{ij} P = \tau_{ij}= k |i-j|
\end{equation}
for all $i$ and $j$, where $m_{ij}$ are arbitrary integer numbers. The
symmetry of the delays, $\tau_{ij}=\tau_{ji}$, implies that 

\begin{equation} \label{fase1}
\phi_i - \phi_j +m_{ij} P =  \phi_j - \phi_i +m_{ji} P .
\end{equation}
Thus, the phase differences $\phi_i - \phi_j$ cannot be arbitrary but
have to be either $\phi_i - \phi_j=n_{ij}P$, or $\phi_i - \phi_j=P/2 +
n_{ij}P$, with $n_{ij}$ an integer number.

We shall refer to solutions with $\mod(\phi_i - \phi_j,P)=0$ $\forall$
$i$ and $j$ as {\it in-phase} solutions, and solutions with
$\mod(\phi_{i+1} - \phi_i,P)=P/2$ $\forall$ $i$ as {\it anti-phase}
solutions.  Since $\mod(\phi_{i+1} - \phi_i,P)$ is an integer number,
the period $P$ of the orbit for the anti-phase solution has to be
even. The in-phase and anti-phase solutions verify Eq.~(\ref{fase})
only for certain delay times. For the in-phase solution,
\begin{equation}
\mod(\phi_i - \phi_j,P)=\mod(k|i-j|,P)=0
\end{equation}
$\forall$ $i$ and $j$ only if $k=n P$ with $n$ an integer number; for
the anti-phase solution,
\begin{equation}
\mod(\phi_{i+1} - \phi_i,P)=\mod(k,P)=P/2
\end{equation} 
only if $k=P/2+ n P$ with $n$ an integer number. 

The existence of these globally synchronized states is independent of
the coupling strength; the only requirement is that the periodic orbit
is a solution (stable or unstable) of the logistic map.

To analyze the stability of the globally synchronized solutions we
turn the delayed Eq.~(\ref{mapa}) into a non-delayed equation by the
introduction of auxiliary variables:
\begin{equation}
y_{im}(t)=x_i(t-m)
\end{equation}
where $1\leq i \leq N$ and $0 \leq m \leq M$ with $M=\max( \tau_{ij})$.
In terms of these new variables Eq. (2) becomes
\begin{equation}
\label{nueva}
y_{im}(t+1)= \begin{cases}
              y_{i,m-1}(t)& \text{si $m \ne 0$},\\
              (1-\epsilon)f[y_{i0}(t)]+\frac \epsilon N
              \sum_{j=1}^N f[y_{j,k|j-i|}] & \text{si $m=0$}.
              \end{cases}
\end{equation}
Next we define the vector 
\begin{equation}
Z=(y_{10},y_{20},\dots,y_{N0};y_{11},y_{21},
\dots,y_{N1};\dots ;y_{1M}, y_{2M}\dots,y_{NM})
\end{equation}
which has $N(M+1)$ components. The anti-phase solutions
of period 2 can be written as
\begin{equation}
Z_A^1=(x_a,x_b,\dots;x_b,x_a,\dots), 
\text{  } Z_A^2=(x_b,x_a,\dots;x_a,x_b,\dots)
\end{equation}
and the in-phase solutions of period 2 as
\begin{equation}
Z_I^1=(x_a,x_a,\dots;x_b,x_b,\dots), 
\text{  } Z_I^2=(x_b,x_b,\dots;x_a,x_a,\dots)
\end{equation}
where $x_a$ and $x_b$ are the points of the period 2 orbit of the
logistic map. We re-write Eq. (7) as
\begin{equation}
z_i(t+1) = F_i[z_1(t), \dots, z_{N(M+1)}(t)]
\end{equation}

The in-phase and anti-phase solutions are fixed points of $F^2$:
\begin{equation}
F(F(Z_{I,A}^{1,2}))=F(Z_{I,A}^{2,1})=Z_{I,A}^{1,2}.  
\end{equation}
To analyze the stability of these
solutions we need to calculate the eigenvalues of the matrix
\begin{equation}
A_{ij}={{\partial F_{i}}\over{\partial z_{k}}}\Big|_{Z=Z_{I,A}^2}
{{\partial F_{k}}\over{\partial z_{j}}}\Big|_{Z=Z_{I,A}^1}
\label{matrix}
\end{equation}
This matrix has $N^2(M+1)^2$ elements which are either 1, 0,
$[1-(N-1)\epsilon/N]f'(x_{a,b})$ or $\epsilon/Nf'(x_{a,b})$. Even for
a small array size there is a large number of variables involved and
the eigenvalues have to be calculated numerically.  We observe that
each matrix in the r.h.s of the last equation has $(M+1) \times (M+1)$
blcks with dimensions $N\times N$. Denoting these blocks as
$\mathcal{F}_{ij}$, with $i,j=0,...M$, it is easy to see using
Eq.~\ref{nueva} that the all blocks $\mathcal{F}_{ij}$ with $i>0$ have
all entries $0$, except the blocks $\mathcal{F}_{i+1,i}$ which are the
$N\times N$ identity, $I_N$.  We also note that $\mathcal{F}_{00}=
[1-(N-1)\epsilon/N]f'(x_{a,b}) I_N$ while the  $\mathcal{F}_{0j}$
are nondiagonal, for example  $\mathcal{F}_{01}$ is
\begin{equation}
\begin{pmatrix}
0 & \epsilon/Nf'(x_{a,b})  & \hdots & 0 \\
 \epsilon/Nf'(x_{b,a}) & 0 & \hdots & 0 \\
\vdots & & &\vdots \\
0 &   \hdots &0 & \epsilon/Nf'(x_{a,b}) \\
0 &   \hdots & \epsilon/Nf'(x_{b,a}) & 0 
\end{pmatrix}.
\end{equation}

\section {Results}

In this section we present numerical simulations and results of the
stability analysis, that demonstrate global synchronization in the
in-phase and anti-phase solutions discussed in the previous section.
The stability analysis can only be done for small arrays and small
delay times, since the size of the matrix $A$ [Eq. (13)] increases as
$kN^2$. For large arrays and/or large delays, we simulate the equation
~(\ref{mapa}).  To solve the delay equation ~(\ref{mapa}) we need to
specify the evolution of $x_i(t)$ at times $1\leq t \leq
\max(\tau_{ij})$. We evaluated this by taking $x_i(1)$ a random number
ranging from 0 to 1 and by letting the array evolve initially without
coupling.

\begin{figure}[ht]
\includegraphics[height=6cm]{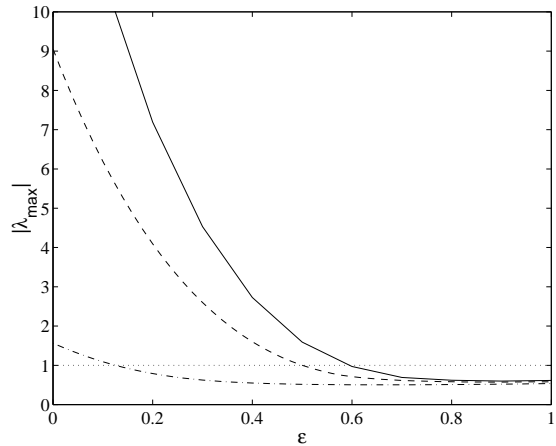}
\caption{Stability analysis of the anti-phase solution of period 2 for 
$k=1$ and $N=12$. We plot the largest eigenvalue of the matrix $A$ 
[Eq. (13)] vs. the coupling strength for $a=$3.5 (dot-dashed line), 
$a=$3.83 (dashed line), and $a=4$ (solid line).}
\label{maxeps}
\end{figure}

First, we show results for the anti-phase solution, which exists for
$k$ even.  For $k=1$, we find that for all values of $a$ there is a
value of $\epsilon$ above which the anti-phase solution of period 2 is
stable. Figure \ref{maxeps} shows the absolute value of the maximum
eigenvalue, $|\lambda_{max}|$, as a function of $\epsilon$, for an
array of $N=12$ maps and three different values of the parameter $a$.
For $a=$3.5 (dot-dashed line) the maps without coupling evolve in a
limit cycle of period 4, for $a=$3.83 (dashed line) the maps without
coupling evolve in a limit cycle of period 3, and for $a=4$ (solid
line) the maps without coupling are chaotic. For clarity the dotted
line indicates the stability boundary $|\lambda_{max}|=1$. In the
three cases, for large enough coupling the anti-phase solution of
period 2 is stable ($|\lambda_{max}|<1$).  Notice that the coupling
strength above which the solution is stable increases with $a$.

\begin{figure}[ht]
\includegraphics[height=6cm]{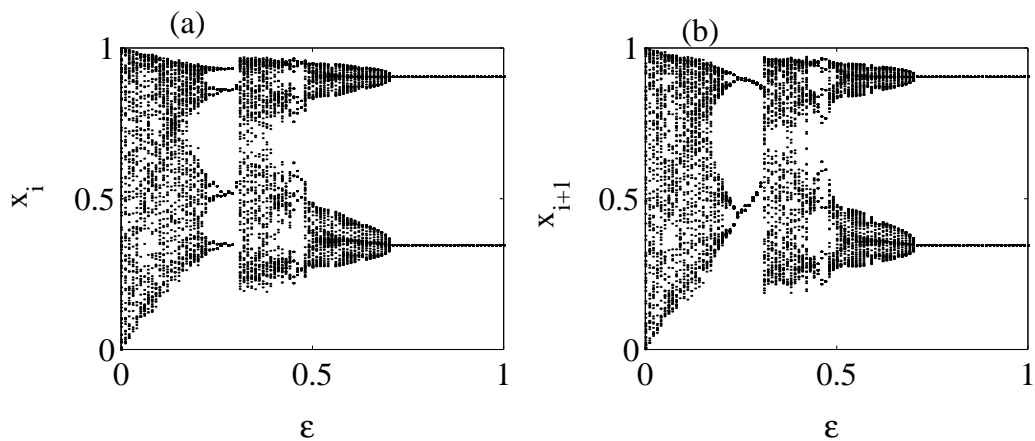}
\caption{Bifurcation diagram obtained numerically, integrating Eq.(2) 
  with $a=4$, $k=1$, and $N=50$. We plot the values of two consecutive
  elements, $x_i$ (a) and $x_{i+1}$ (b). Notice that after a complex
  bifurcation scenario the two elements of the array synchronize in a
  period 2 orbit.}
\label{bif}
\end{figure}

\begin{figure}[ht]
\includegraphics[height=6cm]{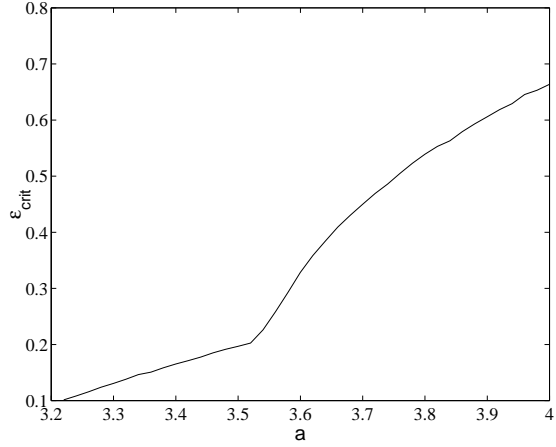}
\caption{Critical coupling strength above which synchronization occurs vs. 
  the nonlinear parameter $a$. $k=1$ and $N=100$.  $\epsilon_{crit}$
  was calculated averaging over 100 different initial conditions.}
\label{tres}
\end{figure}

\begin{figure}[ht]
\includegraphics[height=6cm]{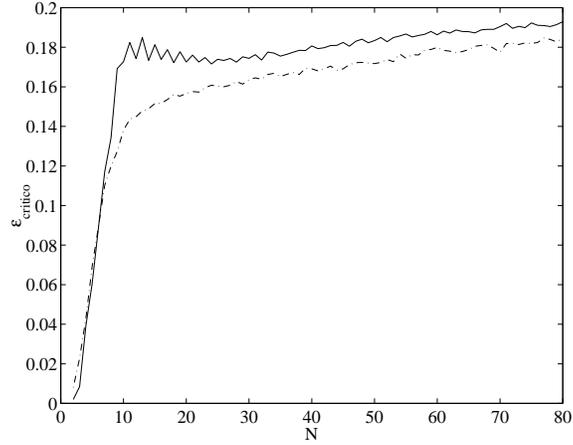}
\caption{Critical coupling strength above which global 
synchronization occurs as 
  a function of $N$ for $a=3.5$ and $k=1$ (solid line); $a=3.5$ and
  $k=4$ (dashed line).}
\label{eigen}
\end{figure}

We verified numerically that for larger arrays the anti-phase solution
is stable.  Figure \ref{bif}(a) displays, as an example, a bifurcation
diagram for $N=50$ ($a=4$ and $k=1$). The bifurcation diagram is done
in the following way: we chose the same initial condition for all
values of $\epsilon$, and we plot the 100 time-consecutive values
$x_i(t)$ (with $t$ large enough) for a given element $i$ of the array.
Figure \ref{bif}(b) displays the same but for a neighboring element.
Above a certain coupling strength the array synchronizes in the period
2 orbit of the uncoupled map and the bifurcation diagram for the two
elements coincide.  The synchronization in the period 2 orbit is
surprising since for $a=4$ the period 2 orbit is unstable for the
uncoupled maps. While the anti-phase solution is stable for $\epsilon
\ge 0.6$ (see Fig. \ref{maxeps}), Fig. \ref{bif} shows that the array
synchronizes in this solution for a slightly larger coupling strength
($\epsilon \sim 0.7$). The critical coupling strength,
$\epsilon_{crit}$, above which global synchronization occurs depends
slightly on the initial condition, and increases with $a$ and $N$.
Figure \ref{tres} displays the critical value of $\epsilon$
(calculated averaging over $100$ different initial conditions) vs.
$a$.  $\epsilon_{crit}$ increases linearly in the parameter region
where the uncoupled maps are periodic, and abruptly in the parameter
region where the uncoupled maps are chaotic.  Figure \ref{eigen}
(solid line) shows that $\epsilon_{crit}$ also increases with the system
size $N$.

Notice that below the critical coupling strength, $\epsilon_{crit}$,
the bifurcation diagrams shown in Figs. \ref{bif}(a) and \ref{bif}(b)
differ. This is due to the fact that for $\epsilon < \epsilon_{crit}$
the array splits into a complex clustered structure. The clustering
behavior in the simpler case when the uncoupled maps evolve in a
period 2 orbit was studied in \cite{Physica_A}.

\begin{figure}[ht]
\includegraphics[height=6cm]{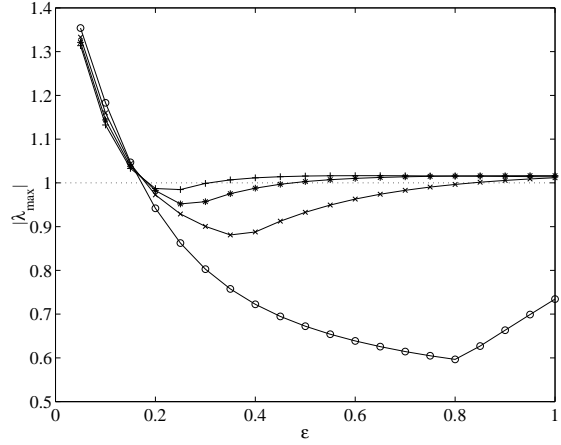}
\caption{Modulus of the largest eigenvalue of the matrix $A$ as 
  a function of $\epsilon$ for the antiphase solution and $a=3.5$,
  $k=1$ (o), $k=3$ (x), $k=5$ (*), and $k=7$ (+).}
\label{curvas}
\end{figure}

\begin{figure}[ht]
\includegraphics[height=8cm]{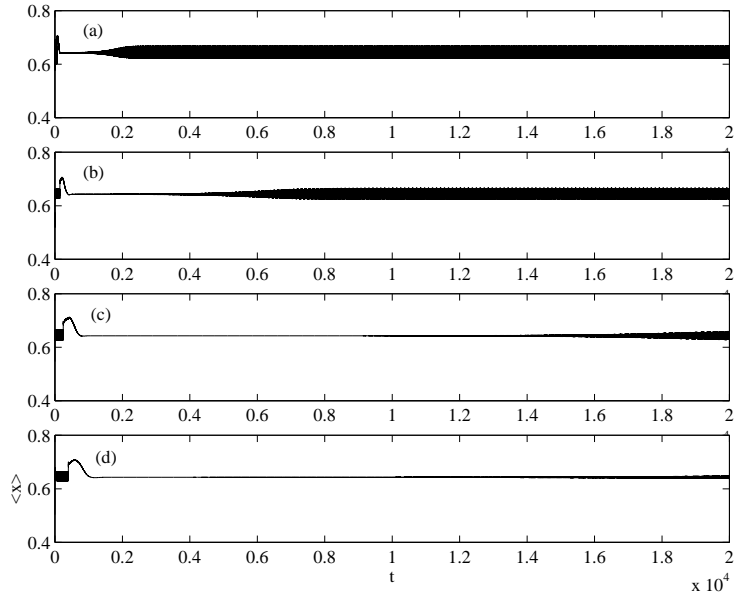}
\caption{Temporal evolution of the mean value $<x_i(t)>$ for four different
  system sizes, $N=12$ (a), $N=30$ (b), $N=50$ (c) and $N=80$ (d). The
  parameters are $a=3.5$, $k=5$, and $\epsilon=0.6$}
\label{xx}
\end{figure}

For larger time delays and $k$ odd, the interval of coupling strength
in which the anti-phase solution of period 2 is stable becomes more
narrow. As an example, Fig. \ref{curvas} displays $|\lambda_{max}|$ vs
$\epsilon$ for $a=3.5$ and $k=1,3,5,$ and $7$. Note that in a wide
range of coupling strength $|\lambda_{max}|$ is slightly larger than
1. In this parameter region, starting from random initial conditions
there is a transient time in which the array approaches the anti-phase
solution; after this transient the array exhibits a complex
spatio-temporal behavior. The transient time increases with $N$; as an
example, Fig. \ref{xx} displays the mean value, $<x>=\sum_{i=1}^N x_i$,
vs.  time, for four different system sizes. The study of this
unexpected effect of the system size is the object of future work.

\begin{figure}[ht]
\includegraphics[height=6cm]{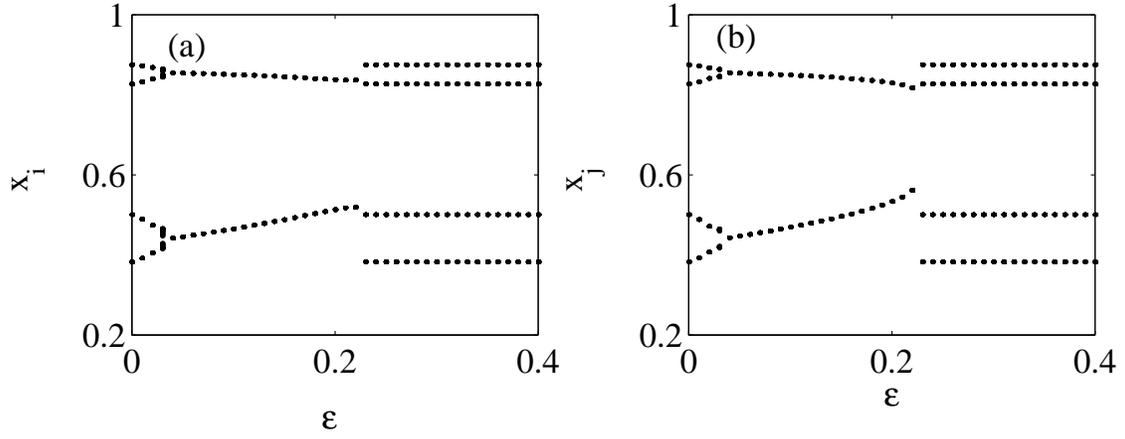}
\caption{Bifurcation diagram obtained numerically, integrating Eq.(2) 
with $a=3.5$, $k=4$, and $N=50$. 
We plot the values of two different elements of the array, 
$x_i$ (a) and $x_j$ (b). Notice that after a period-halving bifurcation 
the elements of the array synchronize in a period 4 orbit.}
\label{bif2}
\end{figure}

Next, we show results for the in-phase solution, which exists for
$k=nP$. Figure \ref{bif2} displays the bifurcation diagram for two
elements of the array, and $a=3.5$, $k=4$, and $N=50$ (for $a=3.5$ the
uncoupled maps evolve in a orbit of period 4). We observe that above a
critical coupling strength ($\epsilon_{crit}\sim 0.23$) the array
synchronizes in the period 4 orbit of the uncoupled map. As in the
case of the anti-phase solution, for coupling strengths below
$\epsilon_{crit}$ the bifurcation diagrams shown in Figs.
\ref{bif2}(a) and \ref{bif2}(b), differ. This is due to the fact that
the two elements belong to different clusters.  The dashed line in
Fig. \ref{eigen} shows that $\epsilon_{crit}$ increases with the
system size $N$.

For arbitrary values of $k$, $a$, and $\epsilon$ we found a rich
variety of complex spatiotemporal behaviors. The characterization of
the different dynamic regimes is the object of future work.

\section{Summary and conclusions}

We studied the synchronization of a linear array of identical logistic
maps. We consider time-delayed mutual coupling with delay times
$\tau_{ij}$ that are proportional to the distance between the maps
($\tau_{ij}=k|i-j|$). Depending on the time delays and on the coupling
strength, different synchronization regimes might occur. If the
coupling is weak the array usually splits into a complex clustered
structure. If the coupling is large enough, global synchronization
occurs. In the globally synchronized state each element of the array
sees all other elements in its present state [$x_i(t)=x_j(t-\tau_{ij})
\forall i,j$], and all the elements of the array evolve along a
periodic orbit of the uncoupled maps. The spatial correlation along
the array is either periodic or homogeneous depending on $k$. If $k$
is odd the array synchronizes in anti-phase, such that the state at
time $t$ of two consecutive elements is $x_i(t)=x_0(t)$,
$x_{i+1}(t)=x_0(t+P/2)$ (where $x_0(t)$ is a particular realization of
the orbit of period $P$, used as a reference). If $k=nP$ the array
synchronizes in in-phase, such that the state at time $t$ is
$x_i(t)=x_0(t)$ $\forall i$.  For parameter values such that the
uncoupled maps are chaotic, mutual delayed coupling suppresses chaos
rendering the evolution of the elements of the array periodic in time.
Thus, an important consequence of our analysis is that delayed
coupling might allow controlling a chaotic array by rending an
unstable periodic orbit of the uncoupled maps, stable.

\end{document}